\documentclass[journal]{IEEEtran}
\ifCLASSINFOpdf
  \usepackage[pdftex]{graphicx}
\else
\fi
\usepackage{comment}

\usepackage{amsmath}
\usepackage[dvipsnames]{xcolor}

\begin{document}
%
\title{CubeSounder: Low SWaP-C 180 GHz Radiometer for Atmospheric Sensing Tested on High Altitude Balloons}
%
%
%

\author{Kyle D. Massingill, Tyler M. Karasinski, Sean Bryan, Michael Baricuatro, Daniel Bliss, Delondrae Carter, Walter Goodwin, Jonathan Greenfield, Christopher Groppi, Philip Mauskopf, Philip Rybak, Scott Smas, Roshni Suresh, Sage Tinlin, Bianca Wullen, and Peter Wullen
\thanks{K. Massingill is with the National Radio Astronomy Observatory,  Socorro,
NM, 87801-0387 USA e-mail: kmassing@nrao.edu.}
\thanks{T. Karasinski, S. Bryan, D. Carter, W. Goodwin, J. Greenfield, C. Groppi, P. Mauskopf, B. Pina, P. Rybak, S. Smas, S. Tinlin, and P. Wullen are with the Arizona State University School of Earth and Space Exploration.}
\thanks{M. Baricuatro, D. Bliss, and R. Suresh are with the Arizona State University School of Electrical, Computer, and Energy Engineering.}

\thanks{Manuscript received March, 2026.}}

%
%

\markboth{Submitted to IEEE TRANSACTIONS ON INSTRUMENTATION AND MEASUREMENT}%
{Shell \MakeLowercase{\textit{et al.}}: Bare Demo of IEEEtran.cls for IEEE Journals}
%



\maketitle

\begin{abstract}
Microwave sounding is the leading driver of global numerical weather forecasting, but is limited by the scalability of such instruments. With modern machining and commercial microwave components, it is now possible to design low size, weight, power, and cost (SWaP-C) microwave spectrometers while maintaining wide bandwidth performance. Here we report on the status of CubeSounder, a spectrometer tailored for water vapor radiometry that utilizes passive wave guide filter banks. After developing a prototype and high altitude balloon payload, we demonstrated CubeSounder on commercial stratospheric balloon flights. We report on our design process, especially the simulation and fabrication of the custom millimeter-wave filter banks. We also report the initial results of the data collected from the balloon flights.
\end{abstract}

\begin{IEEEkeywords}
millimeter-wave, microwave sounding, water vapor, stratospheric balloons, spectrometer.
\end{IEEEkeywords}

\IEEEpeerreviewmaketitle

\section{\label{sec:level1}Introduction}
\IEEEPARstart{D}{ata} from microwave radiometers on large U.S. weather satellites is the single highest impact driver of global weather forecasting \cite{CardinaliCarla2009Mtoi,NAP24938}. These satellites carry instruments such as the Advanced Technology Microwave Sounder (ATMS) \cite{weng2012,Kim2014} and Advanced Microwave Sounding Unit (AMSU) \cite{536029}. These sensors use a heterodyne mixer followed by RF signal processing systems. This effective approach comes with a relatively high size, weight, power and cost (SWaP-C), as well as system complexity. In addition, this approach requires a frequency-stabilized local oscillator.

These existing weather satellites systems deliver valuable weather data but will be expensive to replace as they reach end of life. Spectroradiometers from astronomy (SuperSpec \cite{Redford2021SuperSpec}, DESHIMA \cite{karatsu2026deshima20200400ghz}, Microspec \cite{Cataldo:14}) are promising but are not mature and require sub-Kelvin cooling that is not suitable for a low-SWaP-C application.

\begin{figure}[h]
\centering
\includegraphics[scale=0.47]{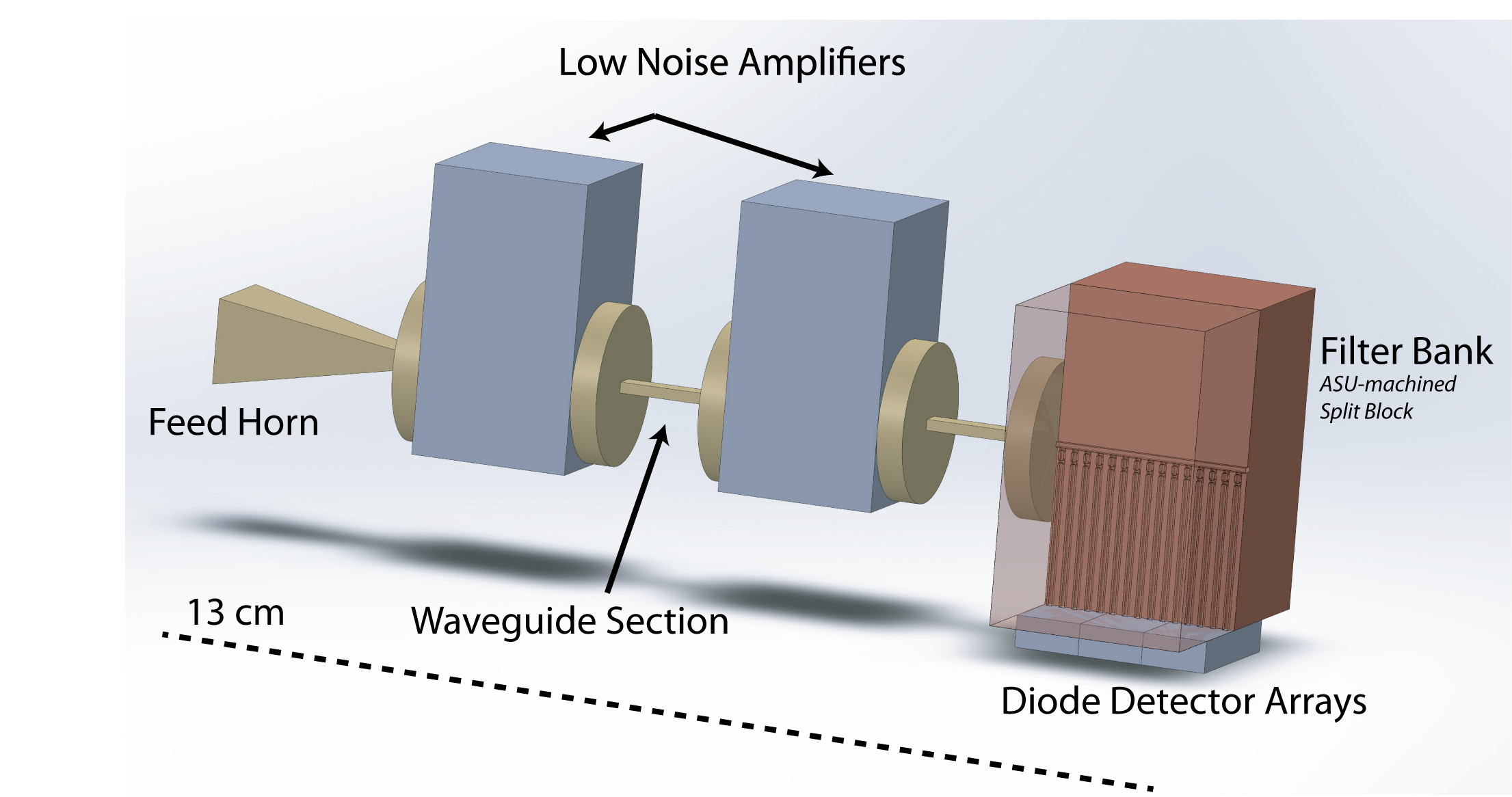}
\caption{Drawing of spectrometer system. Broadband signal enters the feed horn, is then amplified by LNAs and finally divided by filter-bank. The filter-bank is a passive component of directly milled metal based on a novel design.}
\label{fig:spectrometer}
\end{figure}

These disadvantages of existing spectrometer systems led us to develop millimeter-wave waveguide filter bank technology \cite{Bryan2015} at Arizona State University in the CubeSounder program supported by NASA Flight Opportunities. Our approach, as illustrated in the conceptual overview in Fig. \ref{fig:spectrometer}, is to first amplify the signal from the scene with a commercial millimeter-wave low noise amplifier (LNA). The signal is then channelized through a custom designed millimeter-wave filter bank. The filter bank consists of resonant cavities that couple each frequency channel to a different output waveguide port. A millimeter-wave diode power detector at each output port detects the signal which a commercial instrumentation amplifier then amplifies before digitization, and storage.

Some of the authors have described lab testing of our waveguide filter bank technology at 90 GHz \cite{Bryan2015} and 180 GHz \cite{Bryan2016} in the laboratory. In addition, an early publication \cite{massingill2020} by some of the authors gave an overview of CubeSounder mission goals as well as early lab measurements of a single channel lab based prototype.

In this paper, we describe the design, laboratory performance, and successful series of high-altitude balloon flights of CubeSounder. This includes lab testing of our multichannel 183 GHz instrument, as well as the design and fabrication of a multichannel 60 GHz radiometer. We also describe integrating these instruments into a balloon payload, and show results from our series of four high-altitude flights. These flights raised the technical readiness level of CubeSounder technology to TRL6 on the NASA scale.


\section{Instrument Architecture}
CubeSounder is a dualband direct detection spectro-radiometer system operating with passbands near 60 GHz (V band) and 180 GHz (G band). Instead of using a mixer as in conventional systems, we amplify the broadband signal with a broadband LNA, then split the signal into passbands with a filter bank, then directly direct the signal at each filter bank output with millimeter wave detectors.

As illustrated in Fig. \ref{fig:spectrometer}, light enters through a pyramidal feed horn, then is amplified at broadband by two LNAs. Frequencies are then selected off as the light travels through the filter-bank. Each channel of the filter-bank terminates on a diode detector. The LNAs and readout electronics are the only parts of the spectrometer that require power, and the size is mostly limited by the case size of LNAs and diodes. For our 180 GHz system we used the Radiometer Physics 03000033 LNA and Pacific Millimeter Products detectors. For our 60 GHz band we used an Eravant SBL-5436631850-1515-E1 LNA and Pacific Millimeter Products Detectors.

\begin{figure}[ht]
\centering
\includegraphics[scale=0.32]{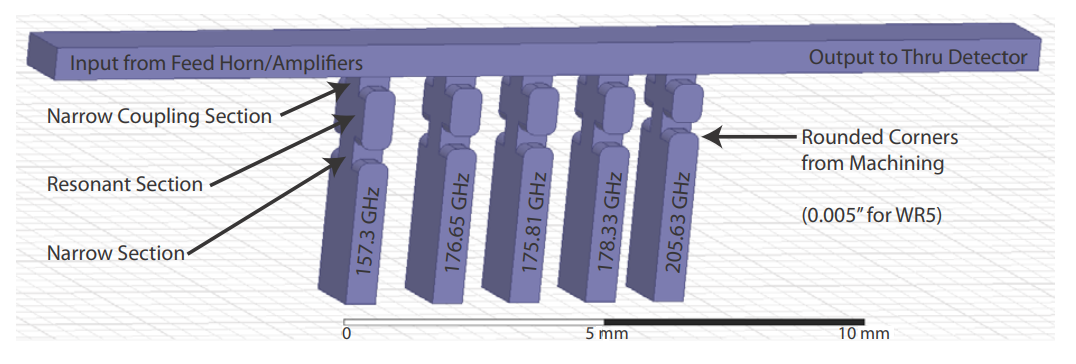}
\caption{An example of the waveguide filter-bank concept. Broadband light propagates through the waveguide and different frequency bands are selected off of the main waveguide by the five channels. Center frequency of each channel is defined by length of resonant cavity.}
\label{fig:Filter bank example}
\end{figure}

\subsection{Filter-bank}

Our filter bank consists of a series of resonant cavity waveguide filters arranged along a main broadband central waveguide (see Fig. \ref{fig:Filter bank example}). We designed the filter bank using batched simulation and scattering matrix (S-matrix) cascading. This reduces the processing needed to simulate the filters and allows for scalability to n-channels. In our approach, we simulated the performance of each individual spectral channel using commercial E\&M simulation software, then combined each channel to simulate the end-to-end performance of the entire spectrometer system. Simulating a 10 channel filter bank can be done in $\sim$hour with a six core processor, which enables us to iterate the design to yield passband centers and channel bandwidths that meet our requirements. The design workflow utilizes \textit{CST studio} for 3D E\&M simulation and a python script utilizing \textit{scikit-rf} (an object oriented radio frequency engineering library) for S-matrix cascading and filter spacing (the physical distance between filters along main waveguide) determination. The original concept for mm-wave filter banks are described in \cite{Bryan2015} and \cite{Bryan2016}.

The individual filters are tuned by a narrow coupling section, then a half-wavelength resonating cavity. A second narrow section, on the other end of the resonant cavity, defines the length of the resonator. The narrow sections have a cutoff frequency 50\% higher than the center frequency of the channel. The center frequency of each channel is defined by the length of the resonant cavity, while the bandwidth is defined by the length of the narrow cutoff sections. There is more waveguide loss per length due to skin effect in the narrow cutoff sections of the filters at center frequency. This lowers the optical efficiency of the filter, but is tolerable (with gain) while achieving bandwidths of $\sim$1 GHz. Achieving a higher spectral resolution would require longer cutoff sections, increasing the loss in the filter. We designed the G-band filter-bank with 2 GHz of half power bandwidth and the V-band with 0.5 GHz. Once the filter dimensions have been defined, each channel is individually simulated as a 3D E\&M object, calculating a 3 port S-matrix. The through power of each individual channel can be predicted from the S-matrix; however, this does not account for the interactions between the channels. To reach a final prediction of the filter bank performance, we cascade the S-matrices of the filters and the connecting waveguide sections.

\begin{figure}
    \centering
    \includegraphics[width=0.85\linewidth]{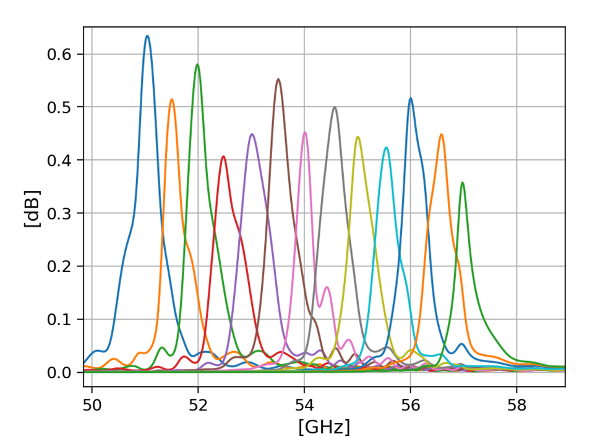}
    \caption{Results from our end-to-end filter bank design and simulation software. Each channel is constructed in CST studio individually then the S-matrices are cascaded with a Python script. Figure shows the predicted through power of the V-band filter bank.}
    \label{fig:60GHz_sim_cascade}
\end{figure}

Between the channels, there are reflections and interference; furthermore, the input of each channel is the output of the previous channel. These interactions and how they ultimately affect the through power of each filter depend on the spacing between the filters. To optimize this spacing for the best overall filter performance (which will differ from optimizing for a single filter), we wrote a python script that allows for quick iteration on the length of the coupling waveguide sections, in terms of number of in-guide wavelengths. The \textit{scikit-rf} library was used for the networking of the S-matrices derived from simulation and standard rectangular waveguide S-matrix for the connecting sections. The result of this cascading and optimizing process can be seen in Fig. \ref{fig:60GHz_sim_cascade} with the calculated through power of the V Band filter-bank.

Once the dimensions and spacing of each filter in the filter bank are defined by the RF simulations, we design the filter banks into a machinable package. We choose to utilize a split-block design where each side of the block has the waveguides milled into it at half depth. Standard rectangular waveguide is used to bring the signal from the in port into the filters and take the signal from the filters to their respective out ports (WR-5 for G-band and WR-15 for V-band). In the split blocks that we have designed so far (G-band and V-band), the filters themselves take up a very small ($\sim$10\% by area) part of the bank. This is because the final block must accommodate the out port of the filters being spaced based on the size of the detectors. For our V-band prototype, we utilized Pacific Millimeter VDH broadband detectors which are packaged in a case $\sim$1x1 inches in size. Therefore, each filter out port was spaced $\sim$1 inch from each other. The spacing of the out port defines the size of the overall block. The split block also includes a thru channel that connects the input signal to the filters, then to a final detector that captures the signal not captured by the filters. 

\begin{figure}
    \centering
    \includegraphics[width=0.85\linewidth]{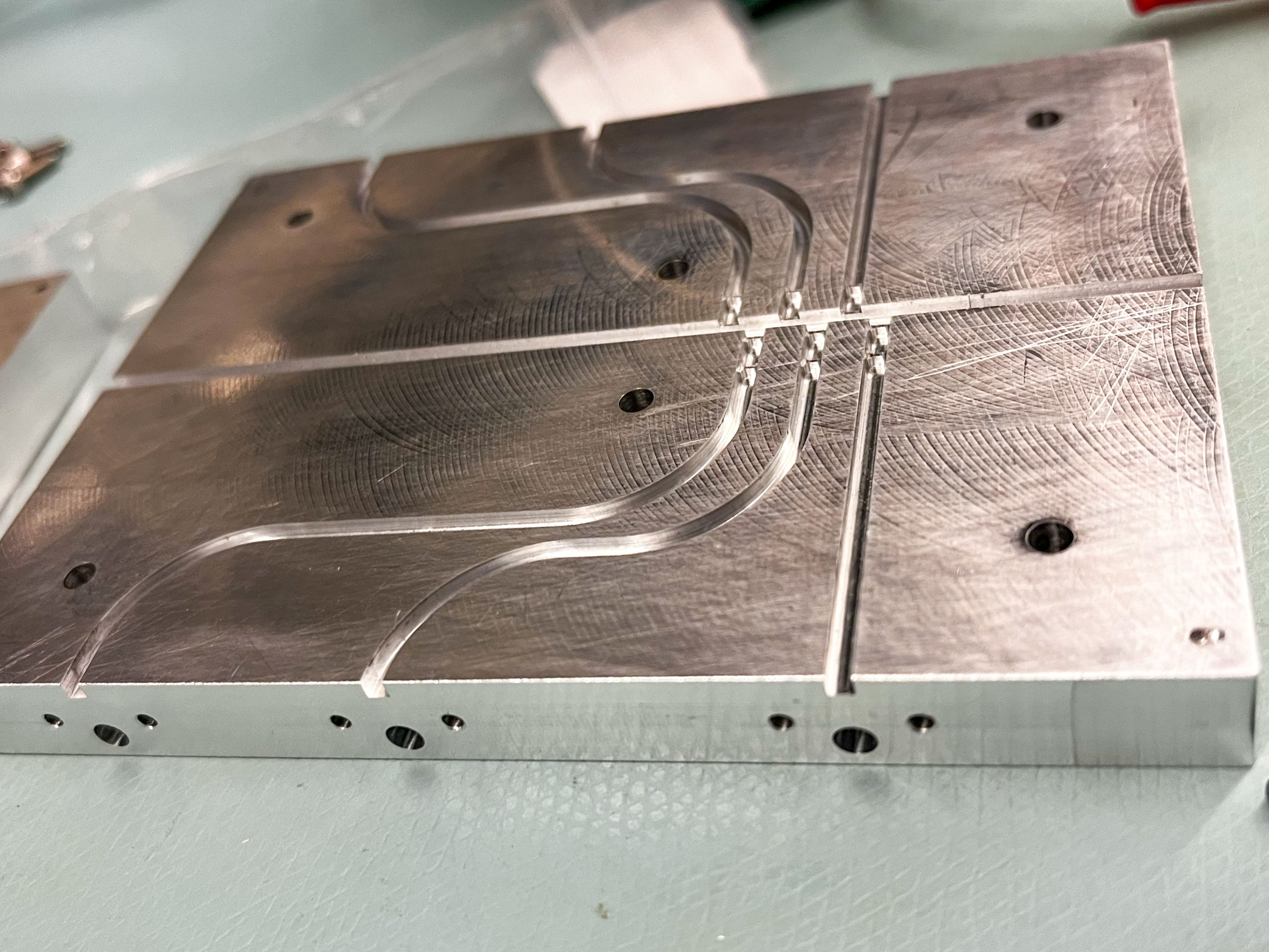}
    \caption[Machined V-band split block]{An aluminum machined split block half from the V-band prototype filter bank. Waveguide sections were milled to half the depth of a standard WR-15 rectangular waveguide. When mated to the other half of split block, the cavity will be the standard waveguide size. Resonant cavity filters split off of main thru channel. Screw and dowel holes are included for split block mating and alignment. This block was machined by Xometry}
    \label{fig:V-band filter milled}
\end{figure}

The G-band prototype was machined in aluminum on a 5-micron tolerance KERN Microtechnik CNC mill at ASU. The V-band prototype was made by Xometry with CNC machining into aluminum with a tolerance of 25-microns (for the sensitive filter areas, other parts of the block were machined to a lesser tolerance). Figure \ref{fig:V-band filter milled} shows half of the completed V-band split block.

\section{Readout Electronics}

Our readout electronics system is significantly simpler than alternate heterodyne architectures. We do not require a high-speed ADC, or real-time FPGA or ASIC processing, to measure the millimeter-wave spectrum. Instead, our readout electronics needs to sample the audio voltage on each of our detectors at a relatively low sample rate, and store the data onboard for later analysis. We used off-the-shelf electronic components readily available from component retailers, and our boards were fabricated on a conventional process by OSH Park and Advanced Circuits.

\begin{figure*}
\centering
\includegraphics[width=0.9\textwidth]{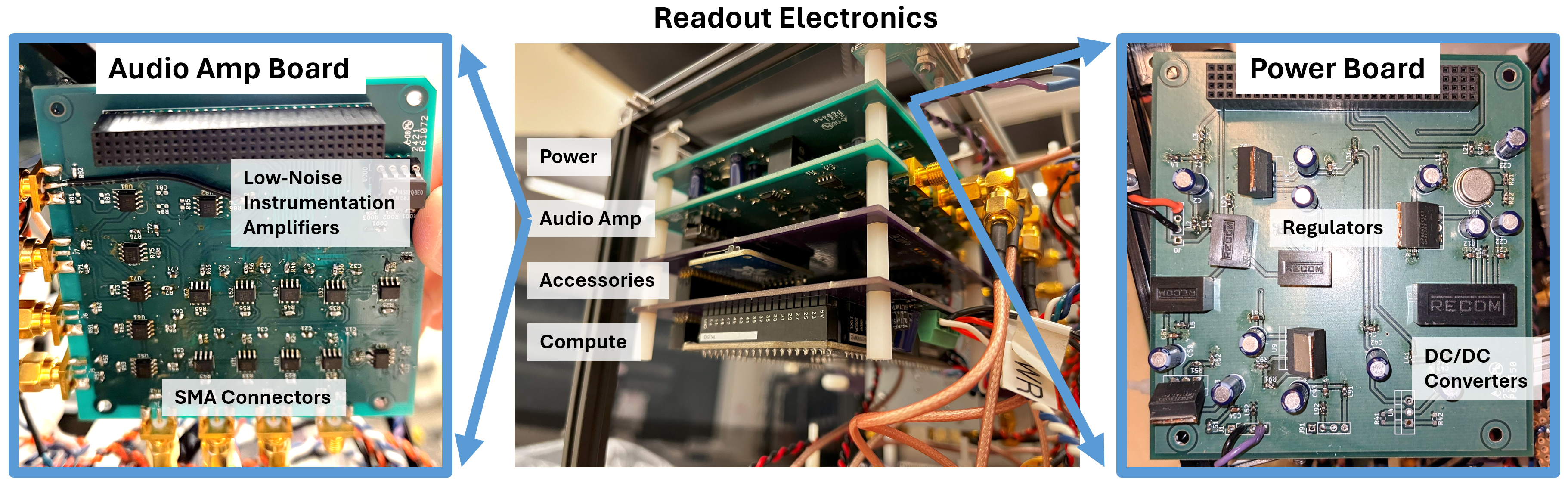}
\caption{The CubeSounder electronics boards which handles readout, storage of sensor data, DC power conversion, as well as the running of accessories such as the real time clock and tiltmeter.}
\label{fig:electronics_boards}
\end{figure*}

For readout, power, and storage, we designed a board stack using a modified version of the PC104 standard. The stack contains four boards (See Figure \ref{fig:electronics_boards}):
\begin{itemize}
    \item Power: Contains DC-DC converters and regulators to provide steady voltage to the rest of board stack, as well as, the spectrometer's LNAs.
    \item Audio Amp: Connects to the filter detectors and amplifies their signal with instrumentation amplifiers.
    \item Accessories: This contains breakout boards for the real time clock (RTC), SD card and tilt meter.
    \item Compute: This contains the Arduino microcontroller and ADC chip.
\end{itemize}

We demonstrate the scalability of CubeSounder by using low-cost and commercially available off-the-shelf (COTS) components for all but custom parts (filters and PCBs being the primary non COTS components). In the CubeSounder spectrometers, each channel of the filter banks are terminated in a diode detector that connects by coax cable to the audio amp board. For both prototypes, we used \textit{Pacific Millimeter Products} (PMP) detectors (VDH detectors for V-band and GD detectors for G-band), the signals from which are amplified by AD8429 audio instrumentation amplifiers. The audio signal is then digitized with an 18 bit analog-to-digital converter (ADC); we use the AD7606C-18 chip. The ADC chip is sampled using an Arduino Mega 2560 microcontroller that runs a script written in C++ for sampling the ADC, RTC and other miscellaneous sensors. The sensor data is converted to plain text and saved to organized files on the SD card on the accessories board. As designed, the audio amplification board only has eight channels. To readout all 14 channels of the V-band spectrometer requires two board stacks.

\section{Radiometer System and Lab Measured Performance}

\begin{figure}
    \centering
    \includegraphics[width=1.0\linewidth]{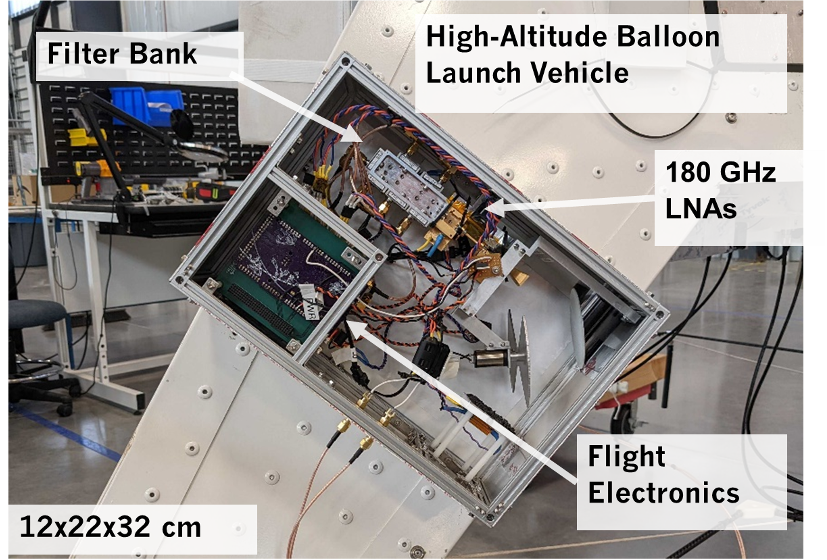}
    \caption{The first iteration of the CubeSounder instrument payload containing the prototype G band spectrometer. The side panel is removed to view internal components. The payload was closed and sealed with aluminum tape before flight.}
    \label{fig:payload1}
\end{figure}

We packaged our CubeSounder prototype radiometers into two different instrument payloads for integration with commercial balloon launch vehicles. The first iteration contained only the G band spectrometer and is pictured in Figure \ref{fig:payload1}. The aluminum enclosure was approximately 12x22x32 cm and included a mirror and chopper wheel in front of the spectrometer feed horn. The chopper switches the spectrometer between looking at the scene and a heated black body pad also in the enclosure. We created a Faraday cage shield around our payload to help reduce RFI/EMI pickup. However, this would block our millimeter-wave signal as well. Thus, we made a millimeter-wave window using a metal mesh filter fabricated on Rogers RO3003 board by Advanced Circuits. This blocked low frequency RF while efficiently passing 60 GHz and 180 GHz.

The second iteration of the CubeSounder payload includes both the G band and V band prototype spectrometers. This payload is larger at 32x22x22 cm and has a mass of 12.5 lbs. The same components from the original G band instrument payload are included, in addition to V band components. For the V band we utilize a separate mirror and switch the spectrometer between a heated thermal load and the scene with an electronic switch. The total power draw of this instrument payload is 17 watts. The flight heritage of these two payloads is discussed in Section V.

\begin{figure}
    \centering
    \includegraphics[width=0.9\linewidth]{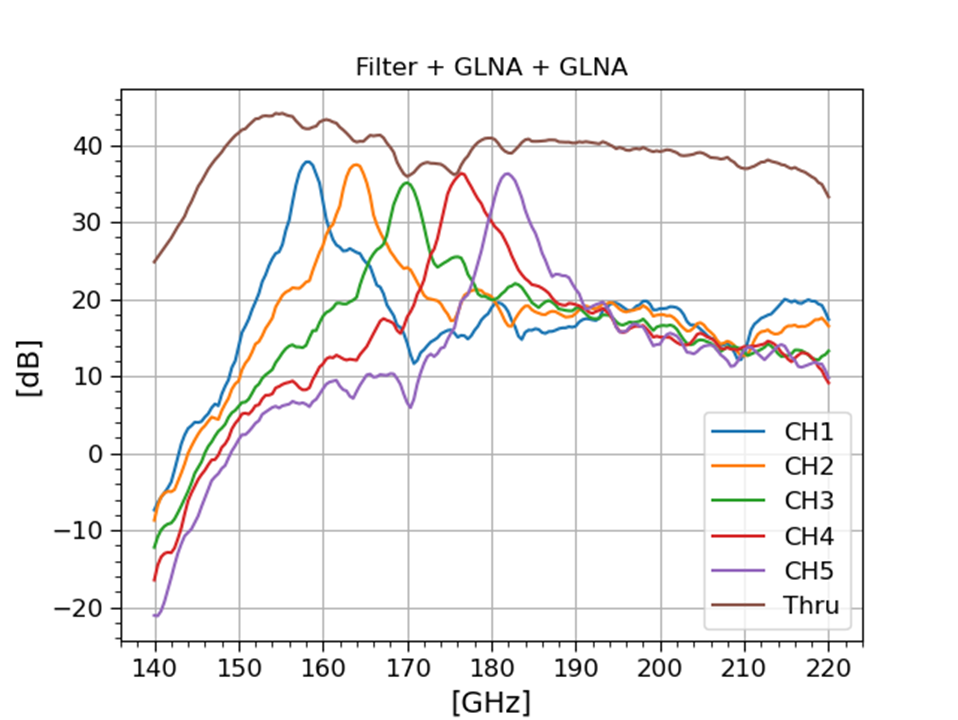}
    \caption[Passbands of G-band Spectrometer]{Passbands of G-band spectrometer prototype determined by measuring the S-matrix of the filter bank with a VNA. The Radiometer Physics G-LNAs were also measured with VNA to determine gain with frequency dependence. The power through the spectrometer is the optical efficiency of the filters plus the gain of the two amplifiers.}
    \label{fig:180 passbands}
\end{figure}

\subsection{G Band Spectrometer}
For the G-band spectrometer, we utilized two Radiometer Physics G-LNAs which provides 20 dB of typical gain (slightly higher toward the center of the band) with a 6 dB noise figure.  The detectors that terminate the filter channels are PMP GD Broadband Detectors with a responsivity of $R=$ 450 mV/mW. The detector noise is estimated by the manufacturer to be $N_{detector}\sim$ 50 pW/$\sqrt{\text{Hz}}$. The signal from the detector goes through an audio amplification stage (AD8429) providing $\sim$34 dB of gain with $N_{Inst.Amp}$ = 1 nV/$\sqrt{\text{Hz}}$ of input noise. Then the signal is digitized using an 18 bit ADC (AD7606C-18).

\begin{figure}
    \centering
    \includegraphics[width=0.9\linewidth]{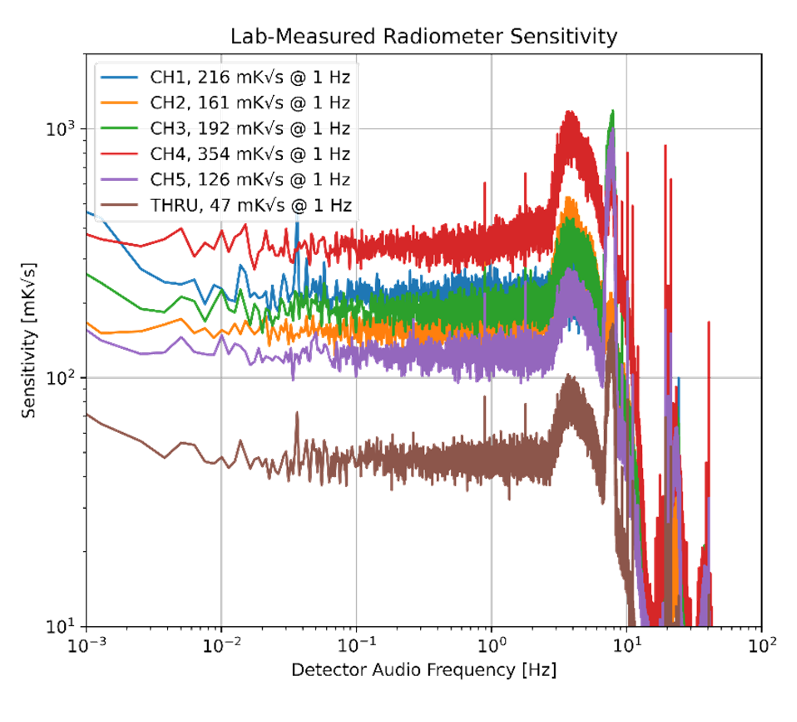}
    \caption[Sensitivity of G-Band Spectrometer]{Sensitivity of 5+1 channel G-band prototype spectrometer. Sensitivity is presented in NET at a sampling rate of 1 Hz. }
    \label{fig:180 Noise Performance}
\end{figure}

The fundamental limit from the radiometer equation is 39 mK$\sqrt{s}$ for per channel sensitivity of the G-band spectrometer. Figure \ref{fig:180 Noise Performance} shows our achieved noise performance per channel of $\sim$200 mK$\sqrt{s}$. Noise propagation shows the noise of the system is dominated by detector noise. The PMP detectors were chosen in part for their low cost and availability. Future prototypes could make use of lower noise Eravant detectors. Noise performance could also be improved by more front end gain, however our review of the commercially available parts indicates the Radiometer Physics LNAs are still the best COTS option in this band.

We measured the S-matrix of the spectrometer with a vector network analyzer (VNA) and G-band waveguide extenders. We utilized a Rhode and Schwarz ZVA 24 with OML 140 to 220 GHz Millimeter Wave VNA Extenders. The extenders act as frequency multipliers and allow us to test via waveguide. For testing, each filter was treated as a 2 port device in the VNA, with the input being the input port of the filter bank (where the LNAs would go in the spectrometer) and the output port being the out port of each individual filter (where the detector would go in the spectrometer). We also measured the S-matrix  of the two G-LNAs. The through power of the spectrometer is presented in Figure \ref{fig:180 passbands}. The passbands shapes of the filter are in agreement with the design. The optical efficiency of the filter is $\sim$20\% and the overall gain of the spectrometer is $\sim$36 dB.

We measured the sensitivity by comparing the response of the spectrometer looking at a room temperature load (293 K) and a load at the boiling temperature of liquid nitrogen (LN2) (77 K). We then map the voltages readout by the ADC to units of temperature using a linear transformation. A linear equation is fit for each channel separately as differences in audio gain and optical efficiency between the filter channels will result in different responsivities. By measuring the standard deviation of each channel in temperature units, we determine the noise equivalent temperature. The NET is calculated as the equivalent noise at a sampling rate of 1 Hz (given in mK$\sqrt{s}$). Figure \ref{fig:180 Noise Performance} shows the measured noise performance. We can compare this to the fundamental limit from the radiometer equation, which for the G-band spectrometer is 26 mK$\sqrt{s}$. This performance limit does not account for sources of noise other than the main amplifier. The measured sensitivity of the channels fall mostly within an order of magnitude of this limit. This shows that the audio amplification is relatively well matched to the detector output and all channels are functional.

\subsection{V Band Spectrometer}

The V-band spectrometer prototype uses an RF switch (Sage Millimeter SKD-5037533035-1515-R1-M) directly after the feed horn which has a nominal insertion loss of 3 dB. This is used in lieu of a chopper to switch between looking at the scene and looking at a heated thermal load. Gain is provided by a single \textit{Eravant} SBL-5037533550-1515-E1 LNA which nominally provides 35 dB of gain in band with a 5 dB noise figure. However, according the manufacturer the LNA has closer to 40 dB of gain from 50-55 GHz, which most of our channels fall into. For the best performing filter channels, we expect optical efficiency of 50-60\% (see Figure \ref{fig:60GHz_sim_cascade}) or $\sim$2.5 dB of loss. 

The PMP VDH detectors have a responsivity of $R=$ 450 mV/mW. The detector noise is estimated by the manufacturer to be $N_{detector}\sim$ 50 pW/$\sqrt{\text{Hz}}$. The instrumentation amplifiers we use, AD8429, have adjustable gain which we have set to $G_{Inst. Amp}\sim$34 dB and $N_{Inst.Amp}$ = 1 nV/$\sqrt{\text{Hz}}$ of input noise. The signal is digitized by an 18-bit ADC (AD7606C-18). 

Testing of the V-band spectrometer is ongoing, but initial in lab sensitivity results show peak channel performance of $\sim$400 mK$\sqrt{s}$. The prototype has yet to successfully perform in flight and we are investigating possible issues especially in the readout part of the system. If gain is not tuned correctly at the readout stage, channels can become saturated under reasonable observing conditions. We are also confirming that component performance matches that reported by manufacturers.

\begin{figure}
    \centering
    \includegraphics[width=1\linewidth]{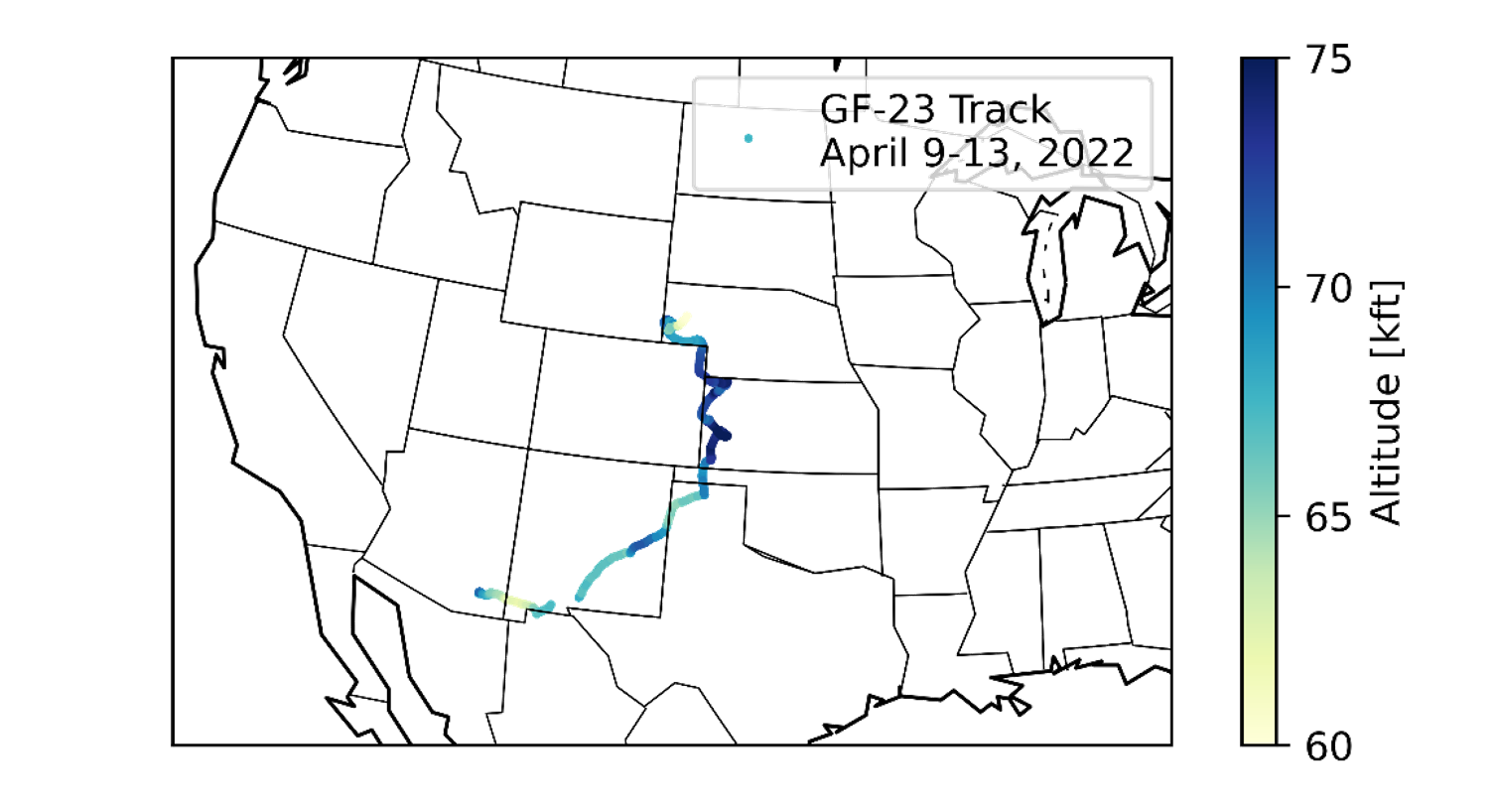}
    \caption{Flight track from April 2022 test flight. This flight carried the first iteration of the CubeSounder instrument payload, pictured in Fig. \ref{fig:payload1}}
    \label{fig:first_flight_track}
\end{figure}

\section{Test Flights}

Supported by the NASA Flight Opportunities program, and in partnership with World View, we conducted a successful series of four high-altitude balloon flights of CubeSounder. After this series, our technology has matured to TRL 6 and we are now finalizing weather data analysis from our final and longest flight.

For our first flight, our goal was to demonstrate integration of our technology onto a World View Stratollite flight vehicle, and attempt to operate our sensor in the flight environment. To do this, we built a minimal test payload that included a 180 GHz low noise amplifier and a minimal set of detector readout electronics. In both lab and pre-flight testing on the vehicle, our radiometer noise performance was not optimal, but we did successfully integrate onto the vehicle and our payload did operate. After launch from the World View facility in Tucson, AZ on October 6th, 2021, the flight unfortunately had to be terminated early. We fully recovered our payload.

In our second flight, our goal was to operate a prototype version of our 180 GHz receiver in flight. For this flight, we had upgraded the detector readout electronics significantly with a new ADC, and added a Faraday cage to improve our isolation from RFI/EMI pickup from the rest of the flight vehicle. We launched from the World View facility in Tucson, AZ on April 9th, 2022 (See Figure \ref{fig:first_flight_track}). We successfully operated in flight, collecting data for four days and maturing our technology to TRL6. Our radiometer noise performance was not as good as we had hoped, but we did detect signals in flight.

For our third flight, our goal was to operate a prototype version of a dualband 60 GHz / 180 GHz receiver in flight. For this version of the payload, we added the 60 GHz band, and continued to improve our rejection of RFI/EMI pickup by analog filtering the power supply and other areas of the system. We launched from the Municipal Airport in Page, AZ on August 16th, 2023. Unfortunately this flight had to be terminated early. We fully recovered our payload.

\begin{figure*}[t]
    \centering
    \includegraphics[width=1.0\linewidth]{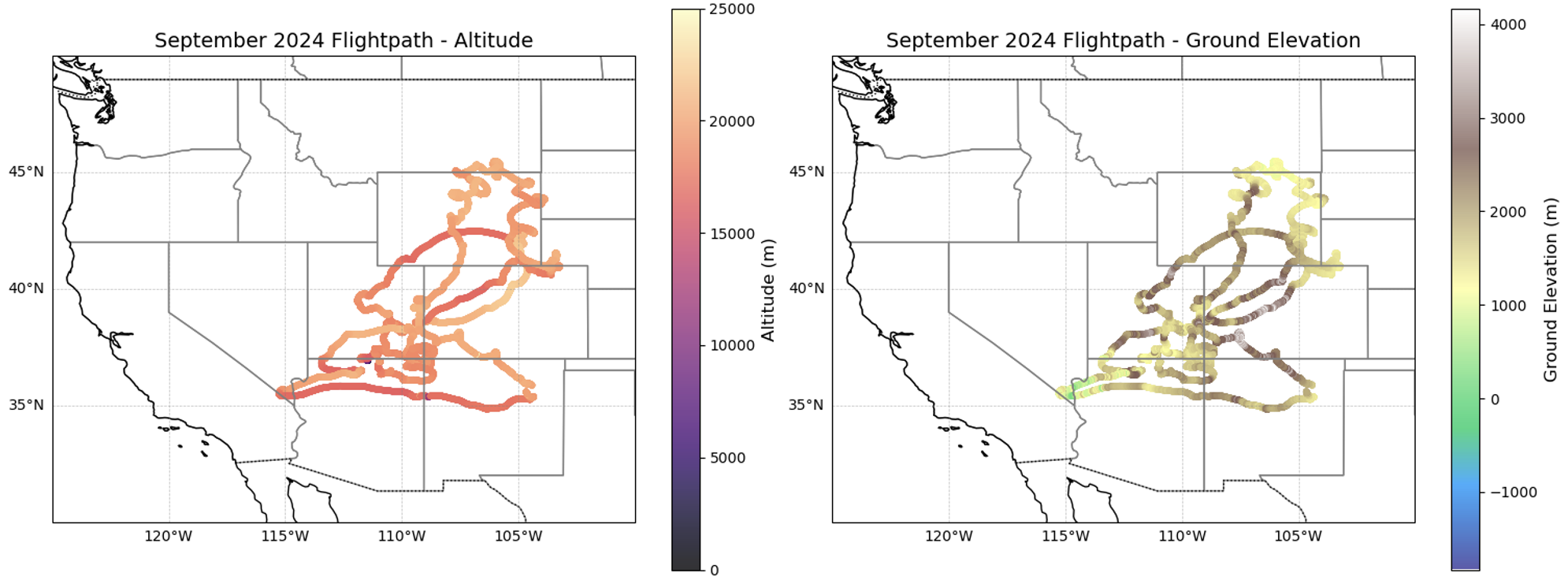}
    \caption{Flightpath for the September 2024 CubeSounder flight aboard the WorldView Stratollite balloon platform, colored according to altitude (left) and ground elevation (right).}
    \label{fig:flightpaths}
\end{figure*}

For our fourth flight, our goals were the same as the previous flight. We continued to improve our RFI/EMI rejection strategy, improving filtering and adding deglitching to our software data analysis. In flight, this yielded noise performance closer to the ground-measured noise shown in Figure \ref{fig:180 Noise Performance}. We launched from the Municipal Airport in Page, AZ on August 31st, 2024. During this successful flight, we obtained a month of high-quality data in operation. The dataset spans from August 31st, 2024 14:00:12 to September 27th, 2024 07:10:07 UTC. As shown in Figure \ref{fig:flightpaths}, this flight spanned stratospheric altitudes up to 32,576.4 m and was bound to the Rocky Mountain region (longitudes 94.47 W to 115.28 W and latitudes 26.49 N to 45.48 N). As shown in the second panel of Figure \ref{fig:flightpaths}, ground elevations over CubeSounder's flightpath ranged from 197 to 4,169 m, derived from NASA's Shuttle Radar Topography Mission 1-arcsecond dataset, distributed via AWS Terrain Tiles (Skadi). This broad range of geographic conditions provides an excellent frame of reference for expected variability in CubeSounder measurements. 

\begin{figure}
    \centering
    \includegraphics[width=1.0\linewidth]{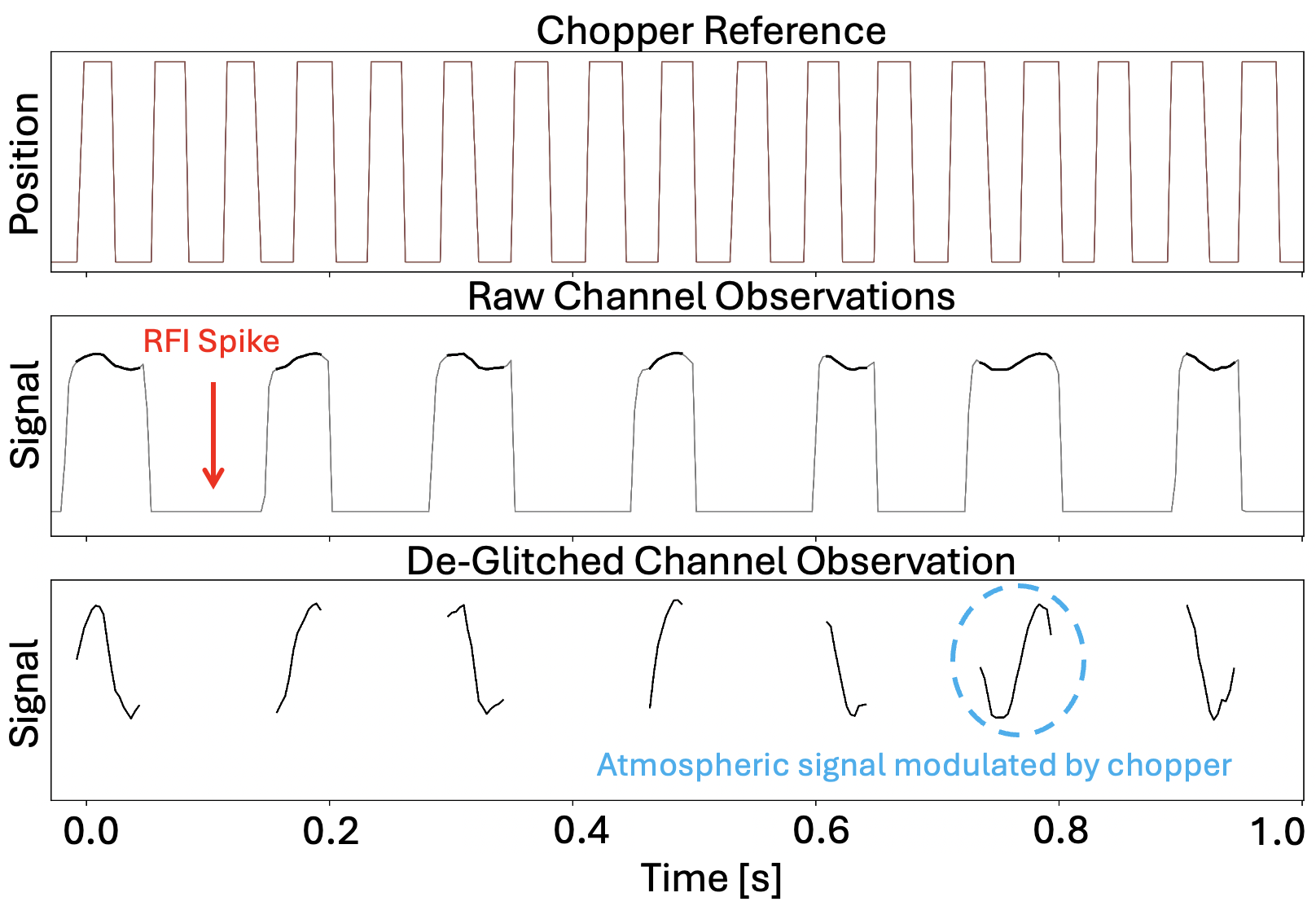}
    \caption{Illustration of single-channel observation from a one-second excerpt of flight data, shown relative to chopper position (top panel) with annotated downward RFI spikes (middle panel) prior to de-glitching. The bottom panel shows the same observation post-de-glitching, with scene data annotated and retained and RFI spikes removed.}
    \label{fig:deglitch}
\end{figure}

As elaborated upon in the following section, data analysis is ongoing of the in-flight noise performance of all three test flights, with particular focus on the complete September 2024 flight.

\section{In-Flight Performance}

The September 2024 balloon flight yielded one month of critical radiometry data reflecting real-world observations in complement to lab-measured performance, providing both real atmospheric data and a test bed for ongoing instrument feasibility investigations.

As expected, real-world atmospheric sounding brings with it unique challenges, including radio frequency interference (RFI) caused by a power supply included on the balloon payload that is not associated with this project. Mitigation efforts during signal processing focus on periodic downward spikes attributed to said power supply, manifesting as overwhelming contamination at magnitudes greater than variations associated with the 183 GHz spectral line. Referred to as `glitches', these appear at regular intervals in time, lending to easy identification when examining flight-measured, per-channel voltage data. Glitches appear as narrow, repeatable spikes, as annotated in figure \ref{fig:deglitch}. As such, simple approaches to data processing are insufficient in removing RFI-associated impacts. Instead, in-flight performance analysis requires a comprehensive workflow, capable of identifying periods of overwhelming RFI which vary in magnitude.

We develop a multi-step data pipeline which mates raw power diode outputs to radiosonde UNIX timestamps, performs robust glitch detection using brute-force, time-based filtering (in which significant glitches are identified relative to a diagnostic median), demodulates signal relative to chopper wheel position (recorded in an independent data column during flight; shown in figure \ref{fig:deglitch}), and maps resulting detector voltages to brightness temperatures via lab-measured $LN_2$ calibration coefficients and in-flight reference load temperature.

Current glitch detection concentrates on isolating extreme contamination from underlying atmospheric variability. We utilize a brute-force approach in which detector voltages are summed across all frequency channels, defining a single diagnostic time series. This is permissible as glitches in-phase across channels. Large deviations from the median of this summed signal are flagged for contamination, and the corresponding time intervals (including a small buffer before and after each gap) are masked. This procedure preserves underlying structure associated with small-scale variation at target frequencies, while mitigating loud periods of RFI. As shown in figure \ref{fig:deglitch}, we find reasonable success with this method. We note, however, there is still considerable work to be done to refine this brute-force approach. Because the applied threshold is rigid and global, it shows suboptimal performance on more subtle interference. While RFI is dramatically minimized, it remains difficult to remove entirely. An investigation into alternative RFI-removal methods is ongoing, including detecting falling and recovering signal edges and RMSE-based thresholds.

Following de-glitching procedures, we demodulate channel measurements relative to mechanical chopper wheel position at a frequency of approximately 17 Hz, isolating variability corresponding to the actual atmospheric scene. This removes long-period gain drift that may otherwise be introduced by instrument components, essentially constituting Dicke-switching (\cite{Dicke1946}). In doing so, we extract the difference in measured voltage between the scene and payload-contained reference thermometer, each measured within one chop cycle, using this delta value for conversion from digitized voltages to absolute brightness temperatures.

To measure brightness temperatures from CubeSounder observations, we use two-point linear calibration. As previously noted, channel responsivity is measured in a controlled lab setting, with room temperature (293 K) acting as hot load, and $LN_2$ (77 K) acting as cold. This provides a ratio of volts-per-degree Kelvin per channel, which is multiplied by demodulated voltages and summed with the reference load temperature to measure final, absolute brightness temperature, providing a consistent, physical baseline for evaluating in-flight instrument performance and 60/183 GHz intensities. This is expressed mathematically as:

\[
T_i = (293 - 77) \, \frac{V_i}{R_i} + T_\mathrm{ref}
\]

Where $T_i$ is absolute brightness temperature, $V_i$ is observed demodulated voltage, $R_i$ is responsivity (lab-measured demodulated voltage), and $T_\mathrm{ref}$ is the temperature as recorded by payload thermometer.

This complete pipeline minimizes transient noise and RFI glitches via filtering and chopper wheel demodulation, producing meaningful brightness temperature measurements via two-point, linear $LN_2$ calibration. The resulting dataset forms the basis for data validation and comparison between the current state-of-the-art and CubeSounder, as well as between in-flight and predicted lab performance. In future work, we aim to explore CubeSounder measurement capabilities relative to the ERA-5 reanalysis dataset via linear regression modeling, further shedding light on instrument outcomes and potential directions for continued improvement.

\section{Conclusions}

CubeSounder is a highly scalable radiometer system now with flight heritage through demonstrations on stratospheric balloons. The technology has reached a TRL of six on the NASA scale. CubeSounder has demonstrated similar levels of sensitivity to the state of the art microwave sounding instruments (such as ATMS) with an order of magnitude lower mass and power. The technology has been demonstrated to be low cost through the use of primarily commercially available components and custom components that are replicable. Full analysis of the flight data will be in a future publication.

The CubeSounder balloon flights are a pathfinder for future testing in space. The technology is well suited for testing on small satellite platforms with its low SWaP-C and data transfer requirements. CubeSounder will be proposed for future technology development grants.


%



\section*{Acknowledgment}
This research was carried out under a contract with the National Aeronautics and Space Administration (80NSSC21K0355). Additional student support was provided by the Arizona State University School of Earth and Space Exploration. It should be declared that this paper, in sections I, II, III and IV reuses some content from dissertation \cite{massingill_Dis} with permission. 

\ifCLASSOPTIONcaptionsoff
  \newpage
\fi



%

\bibliographystyle{IEEEtran}
\bibliography{references}

@article{CardinaliCarla2009Mtoi,
address = {Chichester, UK},
author = {Cardinali, Carla},
issn = {0035-9009},
journal = {Quarterly Journal of the Royal Meteorological Society},
number = {638},
pages = {239--250},
publisher = {John Wiley & Sons, Ltd.},
title = {{Monitoring the observation impact on the short‐range forecast}},
volume = {135},
year = {2009}
}

@book{NAP24938,
address = {Washington, DC},
author = {{National Academies of Sciences}, Engineering and Medicine.},
doi = {10.17226/24938},
isbn = {978-0-309-46757-5},
publisher = {The National Academies Press},
title = {{Thriving on Our Changing Planet: A Decadal Strategy for Earth Observation from Space}},
year = {2018}
}

@article{Bryan2016,
author = {Bryan, Sean and Aguirre, James and Che, George and Doyle, Simon and Flanigan, Daniel and Groppi, Christopher and Johnson, Bradley and Jones, Glenn and Mauskopf, Philip and McCarrick, Heather and Monfardini, Alessandro and Mroczkowski, Tony},
doi = {10.1007/s10909-015-1396-5},
issn = {15737357},
journal = {Journal of Low Temperature Physics},
number = {1-2},
pages = {114--122},
publisher = {Springer US},
title = {{WSPEC: A Waveguide Filter-Bank Focal Plane Array Spectrometer for Millimeter Wave Astronomy and Cosmology}},
volume = {184},
year = {2016}
}

@article{Bryan2015,
author = {Bryan, Sean and Che, George and Groppi, Christopher and Mauskopf, Philip and Underhill, Matthew},
doi = {10.1109/TTHZ.2015.2433919},
issn = {2156342X},
journal = {IEEE Transactions on Terahertz Science and Technology},
number = {4},
pages = {598--604},
publisher = {IEEE},
title = {{A Compact Filter-Bank Waveguide Spectrometer for Millimeter Wavelengths}},
volume = {5},
year = {2015}
}

@article{Dicke1946,
author = {Dicke, R.H.},
doi = {http://dx.doi.org/10.1063/1.1770483},
journal = {Rev. Sci. Instrum.},
pages = {268-275},
title = {{The Measurement of Thermal Radiation at Microwave Frequencies}},
volume = {17},
year = {1946}
}

@article{Redford2021SuperSpec,
	author = {Redford, J. and Barry, P. S. and Bradford, C. and Glenn, J. and Hailey-Dunsheath, S. and Janssen, R. M. and Karkare, K. and LeDuc, H. G. and Mauskopf, P. and McGeehan, R. and Shirokoff, E. and Wheeler, J. and Zmuidzinas, J.},
	journal = {Bulletin of the AAS},
	number = {6},
	year = {2021},
	month = {jun 18},
	publisher = {},
	title = {SuperSpec: Device {Characterization} and {Preparation} for {Telescope} {Deployment} and {Observations}},
	volume = {53},
}

@article{Cataldo:14,
author = {Giuseppe Cataldo and Wen-Ting Hsieh and Wei-Chung Huang and S. Harvey Moseley and Thomas R. Stevenson and Edward J. Wollack},
journal = {Appl. Opt.},
number = {6},
pages = {1094--1102},
publisher = {Optica Publishing Group},
title = {Micro-Spec: an ultracompact, high-sensitivity spectrometer for far-infrared and submillimeter astronomy},
volume = {53},
month = {Feb},
year = {2014},
doi = {10.1364/AO.53.001094},
}

@misc{karatsu2026deshima20200400ghz,
      title={DESHIMA 2.0: A 200-400 GHz Ultra-wideband Integrated Superconducting Spectrometer}, 
      author={K. Karatsu and A. Endo and A. Moerman and S. J. C. Yates and R. Huiting and A. Pascual Laguna and S. Dabironezare and V. Murugesan and D. J. Thoen and B. T. Buijtendorp and S. Cray and K. Fujita and S. Hähnle and S. Hanany and R. Kawabe and K. Kohno and L. H. Marting and T. Matsumura and S. Nakatsubo and L. G. G. Olde Scholtenhuis and T. Oshima and M. Rybak and F. Steenvoorde and R. Takaku and T. Takekoshi and Y. Tamura and A. Taniguchi and P. P. van der Werf and J. J. A. Baselmans},
      year={2026},
      eprint={2601.21603},
      archivePrefix={arXiv},
      primaryClass={astro-ph.IM},
      url={https://arxiv.org/abs/2601.21603}, 
}

@inproceedings{massingill2020,
title = "Millimeter-Wave Filter Bank Spectrometers",
author = "K. Massingill and S. Bryan and C. Groppi and P. Mauskopf and B. Pina and P. Rybak and P. Wullen",
year = "2020",
language = "English (US)",
series = "Proceedings of the 31st Symposium on Space Terahertz Technology, ISSTT 2020",
publisher = "International Symposium on Space Terahertz Technology",
pages = "40--42",
booktitle = "Proceedings of the 31st Symposium on Space Terahertz Technology, ISSTT 2020",
}

@phdthesis{massingill_Dis,
author={K. Massingill},
year={2024},
title={Development of Millimeter-Wave Instruments for Water Vapor Radiometry and Exploring the Regulation of Galaxy Evolution With High-Redshift ALMA Observations},
journal={ProQuest Dissertations and Theses},
pages={139},
isbn={9798383611500},
language={English},
school = {Arizona State University School of Earth and Space Exploration}
}

@ARTICLE{536029,
  author={Mo, T.},
  journal={IEEE Transactions on Microwave Theory and Techniques}, 
  title={Prelaunch calibration of the advanced microwave sounding unit-A for NOAA-K}, 
  year={1996},
  volume={44},
  number={8},
  pages={1460-1469},
  doi={10.1109/22.536029}}

@article{weng2012,
author = {Weng, F. and Zou, X. and Wang, X. and Yang, S. and Goldberg, M. D.},
title = {Introduction to Suomi national polar-orbiting partnership advanced technology microwave sounder for numerical weather prediction and tropical cyclone applications},
journal = {Journal of Geophysical Research: Atmospheres},
volume = {117},
number = {D19},
pages = {},
doi = {https://doi.org/10.1029/2012JD018144},
year = {2012}
}

@article{Kim2014,
author = {Kim, Edward and Lyu, Cheng-Hsuan J. and Anderson, Kent and Vincent Leslie, R. and Blackwell, William J.},
title = {S-NPP ATMS instrument prelaunch and on-orbit performance evaluation},
journal = {Journal of Geophysical Research: Atmospheres},
volume = {119},
number = {9},
pages = {5653-5670},
doi = {https://doi.org/10.1002/2013JD020483},
year = {2014}
}




%








\end{document}